\documentclass[sigconf,nonacm]{acmart}

\usepackage[noend]{algorithmic}
\usepackage{textcomp}
\usepackage{subfig}
\usepackage{algorithmic}
\usepackage{algorithm}
\usepackage{multirow}
\usepackage{diagbox}
\usepackage{pifont} % Add this line to import the 'pifont' package
\usepackage{enumitem}

\AtBeginDocument{%
  }

\setcopyright{acmlicensed}
\copyrightyear{2018}
\acmYear{2018}
\acmDOI{XXXXXXX.XXXXXXX}
\acmConference[Conference acronym 'XX]{Make sure to enter the correct
  conference title from your rights confirmation email}{June 03--05,
  2018}{Woodstock, NY}
\acmISBN{978-1-4503-XXXX-X/2018/06}

\begin{document}

%%
%% The "title" command has an optional parameter,
%% allowing the author to define a "short title" to be used in page headers.
\title{ToolCaching: Towards Efficient Caching for LLM Tool-calling}

%%
%% The "author" command and its associated commands are used to define
%% the authors and their affiliations.
%% Of note is the shared affiliation of the first two authors, and the
%% "authornote" and "authornotemark" commands
%% used to denote shared contribution to the research.

\author{
    % \IEEEauthorblockN{
        Yi Zhai,
        Dian Shen,
        Junzhou Luo,
        Bin Yang
    }
%     \IEEEauthorblockA{
%         \textit{School of Computer Science and Engineering, Southeast University, China}\\
%         Email:
%         \{zhaiyi, dshen, jluo, binyang\}@seu.edu.cn
%     }
% }

% \author{Yi Zhai}
% \author{Junzhou Luo}
% \author{Jianrui Liu}

\affiliation{%
  \institution{School of Computer Science and Engineering, Southeast University}
  \city{Nanjing}
  \country{China}
}

\email{{zhaiyi, dshen, jluo, binyang}@seu.edu.cn}

% \author{Julius P. Kumquat}
% \affiliation{%
%   \institution{The Kumquat Consortium}
%   \city{New York}
%   \country{USA}}
% \email{jpkumquat@consortium.net}

%%
%% By default, the full list of authors will be used in the page
%% headers. Often, this list is too long, and will overlap
%% other information printed in the page headers. This command allows
%% the author to define a more concise list
%% of authors' names for this purpose.
\renewcommand{\shortauthors}{Trovato et al.}

%%
%% The abstract is a short summary of the work to be presented in the
%% article.
\begin{abstract}
Recent advances in Large Language Models (LLMs) have revolutionized web applications, enabling intelligent search, recommendation, and assistant services with natural language interfaces. Tool-calling extends LLMs with the ability to interact with external APIs, greatly enhancing their practical utility. While prior research has improved tool-calling performance by adopting traditional computer systems techniques, such as parallel and asynchronous execution, the challenge of redundant or repeated tool-calling requests remains largely unaddressed. Caching is a classic solution to this problem, but applying it to LLM tool-calling introduces new difficulties due to heterogeneous request semantics, dynamic workloads, and varying freshness requirements, which render conventional cache policies ineffective. To address these issues, we propose ToolCaching, an efficient feature-driven and adaptive caching framework for LLM tool-calling systems. ToolCaching systematically integrates semantic and system-level features to evaluate request cacheability and estimate caching value. At its core, the VAAC algorithm integrates bandit-based admission with value-driven, multi-factor eviction, jointly accounting for request frequency, recency, and caching value. Extensive experiments on synthetic and public tool-calling workloads demonstrate that ToolCaching with VAAC achieves up to 11\% higher cache hit ratios and 34\% lower latency compared to standard policies, effectively accelerating LLM tool-calling in practical applications.
\end{abstract}

%%
%% The code below is generated by the tool at http://dl.acm.org/ccs.cfm.
%% Please copy and paste the code instead of the example below.
%%

% \ccsdesc[500]{Do Not Use This Code~Generate the Correct Terms for Your Paper}
% \ccsdesc[300]{Do Not Use This Code~Generate the Correct Terms for Your Paper}
% \ccsdesc{Do Not Use This Code~Generate the Correct Terms for Your Paper}
% \ccsdesc[100]{Do Not Use This Code~Generate the Correct Terms for Your Paper}

%%
%% Keywords. The author(s) should pick words that accurately describe
%% the work being presented. Separate the keywords with commas.
% %% A "teaser" image appears between the author and affiliation
% %% information and the body of the document, and typically spans the
% %% page.
% \begin{teaserfigure}
%   \includegraphics[width=\textwidth]{sampleteaser}
%   \caption{Seattle Mariners at Spring Training, 2010.}
%   \Description{Enjoying the baseball game from the third-base
%   seats. Ichiro Suzuki preparing to bat.}
%   \label{fig:teaser}
% \end{teaserfigure}

% \received{20 February 2007}
% \received[revised]{12 March 2009}
% \received[accepted]{5 June 2009}

%%
%% This command processes the author and affiliation and title
%% information and builds the first part of the formatted document.
\maketitle

\section{Introduction}

Due to their impressive capabilities, Large Language Models (LLMs) have been widely used in the web ecosystem, not only giving rise to new web applications such as chatbots\cite{ChatGPT,adiwardanaHumanlikeOpenDomainChatbot2020} and code assistants\cite{CursorAICode, GitHubCopilot}, but also empowering traditional web services like search, recommendation, and e-commerce with advanced AI capabilities.  To further extend the functionality of LLMs in these settings, tool-calling\cite{patilGorillaLargeLanguage2024, songRestGPTConnectingLarge2023,schickToolformerLanguageModels2023} has been introduced, allowing LLMs to interact with external tools and APIs. This capability significantly enhances the versatility and utility of LLM-based applications, unlocking their potential to perform complex tasks, such as data retrieval, computation, and manipulation of external systems.

With the advent of tool-calling, increasingly complex tool-chaining workflows have emerged in modern web applications powered by LLMs, where a single model orchestrates multiple external functions in multi-step pipelines to perform sophisticated tasks. For example, ReAct\cite{yaoReActSynergizingReasoning2022} pioneered the integration of reasoning and acting, enabling LLMs to perform sequential tool invocations interleaved with dynamic reasoning steps. To optimize the performance of these workflows, recent works have focused on improving the efficiency of tool-calling execution, such as parallelizing tool calls\cite{kimLlmCompilerParallel2024}, allowing asynchronous tool calls\cite{gimAsynchronousLLMFunction2024}, and partially executing tools concurrently with LLM decoding\cite{xuConveyorEfficientToolaware2024}. 

% These techniques borrow well-established strategies from traditional computer systems, such as parallel execution and asynchronous scheduling, and have shown significant performance gains. However, a common phenomenon observed in LLM inference, redundant or repeated processing of similar requests \cite{kwonEfficientMemoryManagement2023, prabhuVAttentionDynamicMemory2025, yuOrcaDistributedServing2022}, also frequently manifests in tool-calling workflows \cite{singhLLMdCacheImprovingToolAugmented2024}, which these approaches do not fully address. This observation, together with the methodology underlying the LLM KV-Cache mechanism, motivates us to explore another classic yet underexplored optimization strategy inspired by web applications: \textbf{caching the results of tool calls} to eliminate redundant executions and further enhance overall performance. However, \textbf{the unique characteristics of LLMs and their tool-calling mechanisms introduce inherent complexities that traditional caching strategies may not effectively address.}

These techniques borrow strategies from traditional computer systems, such as parallel execution and asynchronous scheduling, and have shown performance gains. However, a common phenomenon observed in LLM inference, redundant or repeated processing of similar requests \cite{kwonEfficientMemoryManagement2023, prabhuVAttentionDynamicMemory2025, yuOrcaDistributedServing2022}, also frequently arises in tool-calling workflows \cite{singhLLMdCacheImprovingToolAugmented2024}, which these approaches do not fully address. This motivates us to revisit another classic optimization strategy widely adopted in web applications: caching the results of tool calls to eliminate redundant executions. Nevertheless, the unique characteristics of LLM tool-calling, such as semantic variations in requests, heterogeneous tool costs, and the need for lightweight integration with inference systems, pose challenges that traditional caching strategies cannot effectively handle.

Firstly, the cacheability of tool calls is highly uncertain: LLMs dynamically generate heterogeneous requests, ranging from information retrieval to state-changing actions, making indiscriminate caching infeasible. Secondly, freshness requirements vary widely; a single TTL policy is either unsafe or inefficient. Thirdly, workloads are highly dynamic: request patterns and popular tools shift rapidly, rendering static strategies ineffective. Hence, robust caching requires adaptive admission and eviction policies that continuously respond to workload characteristics in real time.

\begin{figure}[t]
    \centering
    \includegraphics[width=0.48\textwidth]{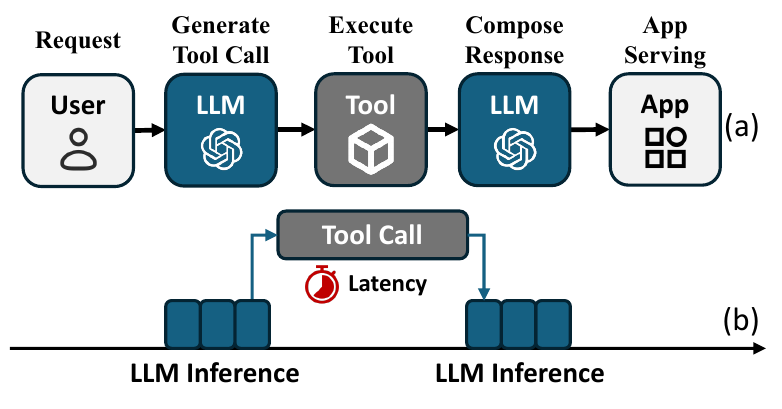}
    \caption{Tool-calling workflow and the corresponding timeline.}
    \label{fig:timeline}
\end{figure}

To tackle these complexities, we recognize two core technical challenges:
\begin{itemize}[leftmargin=*]
    
    \item \textbf{Understanding LLM tool-calling requests.} Tool invocations are highly diverse and context-dependent, making cacheability hard to assess with traditional heuristics. Extracting both semantic and system-level features from dynamic requests is essential yet non-trivial.
    \item \textbf{Designing adaptive caching policies.} Classic criteria such as hit ratio or recency are insufficient for workloads with heterogeneous costs and reuse patterns. Effective policies must integrate richer features to guide admission and eviction decisions.
\end{itemize}

These two challenges form the foundation of our approach. We introduce ToolCaching, a caching framework that integrates request semantics with system-level signals to optimize tool-calling in LLM applications. ToolCaching combines semantic analysis of requests with lightweight observability and in-cache statistics, enabling adaptive admission and eviction policies that are tailored to the unique characteristics of tool-calling workloads. To the best of our knowledge, this is the \textbf{first work to address caching for LLM tool-calling}.

We evaluate ToolCaching on both synthetic datasets and public tool-calling workloads. The results show that ToolCaching improves cache hit ratio by up to 11\% over baseline strategies, and reduces end-to-end latency by up to 34\%, demonstrating its effectiveness in accelerating LLM-based web applications.

Our contributions are summarized as follows: (i) we conduct a comprehensive analysis of the challenges in applying caching to LLM tool-calling and propose \textbf{ToolCaching}, the first framework that integrates semantic and system features for adaptive cache management; (ii) we identify essential features for understanding tool-calling requests and design lightweight mechanisms to collect them; (iii) we develop an efficient cache management algorithm, VAAC, which leverages semantic and system-level signals to guide admission and eviction decisions; (iv) we implement and evaluate ToolCaching on synthetic and public workloads, demonstrating its effectiveness in improving the performance of LLM tool-calling systems.

\section{Background And Motivation}\label{section:background}

This section provides an overview of the background and motivation of our work. We begin by introducing the concept of tool-calling in LLMs, followed by the inherent complexities associated with applying caching in LLM tool-calling scenarios and the importance of cache admission and eviction.

\subsection{Tool-calling in LLMs}

Tool-calling is a mechanism that allows LLMs to interact with external tools and APIs, enabling them to perform tasks beyond their inherent capabilities. This feature has been integrated into various LLMs, such as OpenAI's GPT-4 \cite{achiam2023gpt} and Meta's Llama 2 \cite{touvron2023llama}, allowing these models to retrieve remote data sources, execute computations, and manipulate external systems. Beyond simple single-step invocations, tool-calling also enables the construction of complex multi-step workflows. For example, ReAct\cite{yaoReActSynergizingReasoning2022} demonstrates how LLMs can interleave chain-of-thought reasoning with tool usage, dynamically planning and executing sequences of actions to accomplish sophisticated tasks.

Regardless of task complexity, tool-calling workflows follow a consistent pattern: as shown in \autoref{fig:timeline}a, the LLM parses the user’s request, generates a structured tool call, and sends it to the tool’s API. The returned result is fed back as additional context for response composition, which is finally delivered to the application serving layer.

\begin{table}[b]
\centering
\caption{Representative tool tasks with typical latency and cost.}
\renewcommand{\arraystretch}{1.2} % 调整行间距
\begin{tabular}{|l|c|c|}
\hline
\textbf{Task Type} & \textbf{Latency (ms)} & \textbf{Cost (\$/1k calls)} \\
\hline
Search \cite{GoogleCustomSearch2025} & 700--2000 & $\sim$5.0\\
Wikipedia Fetch \cite{WikimediaAPIs2025} & 200--1000 & 0 \\
Map Planning \cite{GoogleMap} & 50\textendash{}1000 & 5.0 \\
Weather Query \cite{PricingOpenWeatherMap2025} & $\sim$200 & 1.6 \\
% DB Read / Write (DynamoDB, Firestore) & 10--100 & $<$0.001~\cite{aws_dynamo,gcp_firestore} \\
% Email / Message Send (SES, Twilio) & 200--500 & 0.1--8.0~\cite{aws_ses,twilio_sms} \\
% Payment / Transaction (Stripe) & 1000--4000 & $\sim$300.0~\cite{stripe_fee} \\
\hline
\end{tabular}
\label{tab:tool_task_cost_latency}
\end{table}

As depicted in \autoref{fig:timeline}b, tool calls are inserted between two stages of LLM inference, and their execution time can be a dominant contributor to end-to-end latency, particularly when invoking remote APIs over the network \cite{patilGorillaLargeLanguage2024, songRestGPTConnectingLarge2023}. Previous works studies \cite{kimLlmCompilerParallel2024, gimAsynchronousLLMFunction2024,xuConveyorEfficientToolaware2024} focus on optimizing tool-calling performance by adjusting the scheduling between LLM inference and tool execution, enabling parallel processing, and designing more efficient pipelines. 

Besides these execution-level optimizations, our empirical analysis reveals substantial redundancy in tool-calling workloads. In the movie recommendation dataset \cite{srivastava2023beyond, kimLlmCompilerParallel2024}, we found that over 40\% of tool invocations are repeated, indicating significant potential for reusing previous results. Redundancy also frequently arises within a single user session—for instance, a user may first request “top 5 movies,” prompting the system to retrieve and describe a list of titles, and then follow up with “top 5 sci-fi movies,” where overlapping entries lead to duplicate calls for the same movie details. Such repeated invocations not only prolong response time but also amplify the financial and resource costs associated with API usage, since many tools charge per call or consume limited quotas. In \autoref{tab:tool_task_cost_latency}, we summarizes representative tool tasks with their typical latency and cost per 1,000 calls, highlighting the potential benefits of eliminating redundant calls through \textbf{caching the results of tool-calling requests}.

\subsection{Caching in LLM Tool-calling}
Caching has long been an effective strategy in traditional web applications to mitigate redundant computation and network overhead. Recent advances such as the KV-Cache in LLM inference \cite{kwonEfficientMemoryManagement2023, prabhuVAttentionDynamicMemory2025, yuOrcaDistributedServing2022} further demonstrate its power in accelerating system performance. Inspired by these successes, we extend this paradigm to LLM tool-calling: by storing the results of tool invocations in a cache and serving subsequent identical requests directly from cached entries, our approach can substantially reduce repeated tool execution, lower external communication costs, and improve overall end-to-end efficiency.

However, the unique characteristics of LLMs and their tool-calling mechanisms introduce inherent complexities, making the design of caching systems fundamentally different from those in traditional web applications:

\textbf{Uncertain Cacheability of Tool Calls:} LLMs dynamically generate tool calls—ranging from local function executions to remote API invocations—that serve heterogeneous purposes, such as information retrieval or state-changing actions. Some calls require real-time data, while others may trigger irreversible side effects (e.g., sending messages or performing transactions). Caching all tool calls is clearly infeasible: results may rapidly become outdated, and indiscriminate caching of commands can lead to incorrect or unsafe behavior. Accurately determining whether a tool call is safe and beneficial to cache thus demands a deep understanding of each request.

\textbf{Highly Variable Result Freshness:} The freshness requirements of tool-calling results vary dramatically—encyclopedia queries may remain valid for weeks, while financial data or sensor readings may expire in seconds. Applying a uniform cache expiration (TTL) is either wasteful or unsafe. Effectively managing cache staleness requires fine-grained, adaptive TTL assignment that reflects the specific validity of each result, which is especially challenging under diverse and shifting workloads.

\textbf{Rapidly Changing and Unpredictable Workloads:} LLM tool-calling systems routinely face rapidly shifting workloads—request frequencies, popular tools, and user access patterns all fluctuate in response to user behavior or external events. Such volatility makes static cache partitioning and fixed replacement strategies ineffective: what is hot now may soon become cold, and cache resources must be dynamically reallocated. Robust caching management must therefore continuously monitor workload characteristics and adapt cache policies in real time.

\subsection{Caching Management Policy}
Given the characteristics of LLM tool-calling workloads, the design of caching management policy, including admission and eviction mechanisms, plays a crucial role in achieving effective caching. Admission control determines which tool-calling results are admitted into the cache, while eviction policies dictate which entries to replace when capacity is constrained.  

Traditional approaches such as LRU (Least Recently Used) and LFU (Least Frequently Used) have been extensively applied in many systems \cite{einzigerTinyLFUHighlyEfficient2017, frickerVersatileAccurateApproximation2012, zhaoP4LRULRUCache2023}, but they are inadequate for LLM tool-calling scenarios, where the execution cost and latency of each tool call are non-negligible and vary widely. These policies primarily rely on access recency or frequency, without explicitly accounting for the high and uneven cost of tool execution. Recent studies show that adaptive, workload-aware admission and eviction strategies can achieve substantial improvements in hit ratio, resource utilization, and end-to-end performance \cite{guanCACALearningbasedContentaware2019, yangGLCacheGrouplevelLearning2023, yangSegcacheMemoryefficientScalable2021}.

This highlights the need for an effective, workload-aware caching management policy that can adaptively incorporate both the execution cost and latency characteristics of tool calls into its admission and eviction decisions.

\textbf{In response to these complexities and difficulties, drawing inspiration from the recent advances in caching, we propose a novel caching system for LLM tool-calling called ToolCaching.}

\begin{figure}[t]
    \centering
    \includegraphics[width=0.48\textwidth]{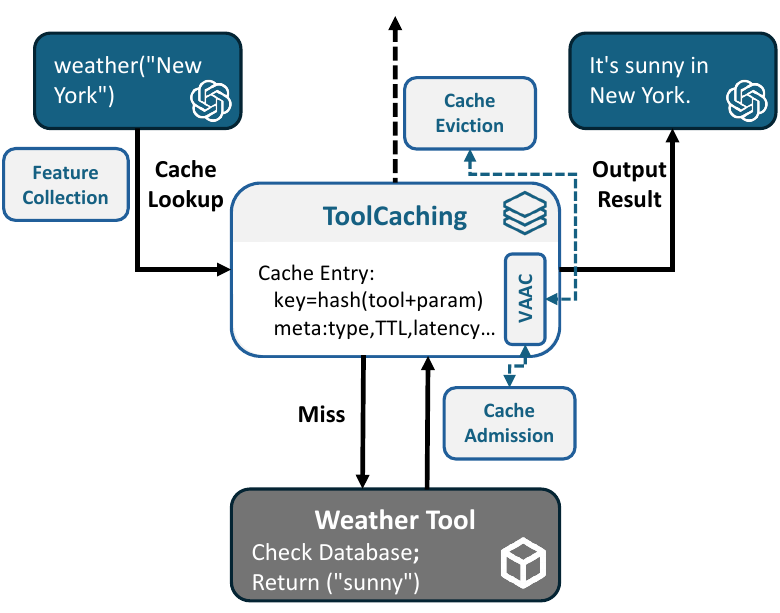}
    \caption{The workflow of ToolCaching.}
    \label{fig:workflow}
\end{figure}

\section{Design}\label{section:design}
In this section, we present the design of ToolCaching. 
As discussed in \autoref{section:background}, caching in LLM tool-calling poses two unique challenges, and to address these issues, ToolCaching is designed around two core components:

\begin{itemize}[leftmargin=*]
    \item \textbf{Feature Collection.} 
    ToolCaching first performs a comprehensive analysis of tool-calling requests by extracting both semantic attributes and system-level indicators. 
    These features, obtained via LLM-based semantic understanding and lightweight system monitoring, provide the foundation for informed caching decisions.
    
    \item \textbf{Caching Management Algorithms.} 
    Based on the collected features, ToolCaching employs an adaptive caching algorithm called VAAC (Value-Aware Adaptive Caching). 
    VAAC integrates two key mechanisms: (i) selective cache admission guided by feature-based grouping and caching value (v-CACA), and (ii) multi-factor eviction (v-LRU) that jointly considers recency, frequency, and execution cost. 
    This design enables ToolCaching to dynamically adapt to heterogeneous and evolving workloads.
\end{itemize}

The workflow of ToolCaching is shown in \autoref{fig:workflow}. ToolCaching intercepts API requests from the LLM. If the request is already in the cache, ToolCaching retrieves the cached result and returns it to the LLM. If the request is new, ToolCaching forwards it to the tool API, extracts semantic features from the request, and determines its cacheability. If deemed cacheable, system-level features are collected and the caching value is evaluated. Based on this value and other features, ToolCaching decides whether to admit the result into the cache or to evict an existing entry. Regardless of whether the result is cached, it is returned to the LLM for further processing. If the request is not cacheable, ToolCaching simply forwards the request to the tool API and returns the result to the LLM.

In the following sections, we will discuss the details of each category and explain how we can overcome the complexities mentioned in \autoref{section:background}.

\section{Feature Collection}\label{section:feature_collection}

Due to the complexities of caching in LLM tool-calling scenarios mentioned above, it is essential to gain a comprehensive understanding of tool-calling requests before designing an effective caching strategy. In this section, we explain which features to collect and how to collect them.

\subsection{Semantic Features}

% \begin{table*}[t]
%     \centering
%     \caption{Selected features for request analyzing and corresponding collection methods.}
%     \renewcommand{\arraystretch}{1.5}
%     \begin{tabular}{|l|l|l|}
%         \hline
%         \textbf{Feature} & \textbf{Description} & \textbf{Collection Method} \\
%         \hline
%         Request Type & INFORMATIONAL (retrieval) or COMMAND (state-changing action) & LLM Semantic Analysis \\
%         Parameter Category & Primary grouping based on input parameters & LLM Semantic Analysis \\
%         Result TTL & Suggested validity duration of tool-calling result & LLM Semantic Analysis \\
%         Associated Users & Set of users who have accessed the cached request & Cache System Statistics \\
%         Request Access Count & Total number of cache hits for the cached entry & Cache System Statistics \\
%         Result Size & Size of the cached tool-calling response & Cache System Statistics \\
%         System Latency & End-to-end latency including network and execution delays & Cache System Statistics \\
%         Observed Resource Cost (Remote) & Billing price for external API calls & Cache System Statistics \\
%         Observed Resource Cost (Local) & CPU consumption for local computations & System-level Monitoring (eBPF) \\
%         \hline
%     \end{tabular}
%     \label{tab:feature-collection}
% \end{table*}

Determining whether a tool-calling request in LLM applications is cacheable is fundamentally a semantic question. Unlike traditional web caching, where cacheability can often be inferred from static resource types or HTTP metadata, LLM tool-calling workloads are driven by dynamically generated context-rich requests with varying degrees of features. We select the following features and explain the reasons.

\textbf{Request Type:}  
Tool-calling extends the capabilities of LLMs by enabling interaction with external APIs that can serve fundamentally different purposes. A single tool may support both \emph{information retrieval} and \emph{command/action} semantics. For example, a messaging tool can be invoked to send a message: 
\texttt{message(send;Mike;"Hello")} or to read the latest message: \texttt{message(read;last message)}.  

This distinction is critical for caching: retrieval-type requests are generally side-effect-free and thus suitable for reuse when repeated, whereas command-type requests often induce irreversible state changes, making their results unsafe or meaningless to cache. We therefore categorize tool-calling requests into two broad classes: \textbf{INFORMATIONAL} (e.g., weather lookup, knowledge base search) and \textbf{COMMAND} (e.g., sending a message, initiating a transaction). To avoid harmful side effects, COMMAND requests are excluded from caching, while INFORMATIONAL requests serve as candidates for reuse. Accurately identifying the request type thus becomes a key feature for determining cacheability and is directly incorporated into ToolCaching's admission control.

\textbf{Parameter Category:}  
This feature captures the grouping of tool-calling requests based on their input parameters. For example, in a weather query tool, requests can be organized by attributes such as location or date (e.g., \texttt{location=London}, \texttt{date=2024-05-01}), which later serve as the basis for request grouping and make caching management adaptive (discussed in \autoref{section:cachemanagement}).

In this work, we adopt a practical heuristic: for multi-parameter tool calls, the first parameter is treated as the primary category for grouping; for single-parameter functions, no further parameter-based grouping is applied. Although we adopt a simple heuristic in this paper, the parameter-based categorization is not fixed; grouping rules can be customized when registering each tool in the cache system to better reflect its semantics and usage patterns.

\textbf{TTL (Time-to-Live):}  
In traditional caching systems, TTL is a fundamental mechanism for controlling data staleness and ensuring the validity of cached content. We adopt the same concept for LLM tool-calling, where TTL specifies the suggested lifetime of a tool-calling result and directly drives cache expiration policies.  

Tool calls with short TTLs (e.g., real-time sensor readings) are less suitable for caching due to rapid staleness, whereas those with longer TTLs (e.g., static knowledge queries) are strong candidates for reuse. In ToolCaching, we define indicative default values: \texttt{0s} for COMMAND requests, \texttt{60s} for real-time data, \texttt{300s} for computational results, and \texttt{3600s} for static knowledge. These values reflect common temporal patterns and can be customized per tool or dynamically adapted to workload characteristics.

Considering all these attributes, the corresponding feature set can be obtained by prompting the LLM with each tool-calling request. An example of the prompting template used for feature collection is shown in \autoref{appendix:prompt}.

In practice, we find that prompt engineering with clear instructions and representative examples, effectively leverages the LLM’s semantic analysis capability and produces accurate and consistent feature extraction results, achieving nearly 90\% accuracy as evaluated in \autoref{tab:llm_semantic_eval}, which forms a solid basis for downstream caching management.

\subsection{System Features}
Once a tool-calling request is identified as cacheable, ToolCaching evaluates the potential benefit of retaining its result by collecting a set of system-level features. These features serve two purposes: (i) estimating the caching value of each entry and (ii) driving adaptive admission and eviction policies under dynamic workloads.

We distinguish two categories of system features:

\emph{In-cache statistics}, maintained directly by the caching system with minimal overhead:  

\textbf{Associated Users:} This feature records the set of users who have accessed each request. In LLM tool-calling scenarios, different users often exhibit distinct patterns of accessing particular tools or queries. By tracking associated users, ToolCaching can effectively group requests based on user interest, as users may repeatedly invoke similar or identical tool calls. For example, a particular user might frequently request personalized movie recommendations, while another consistently queries financial data. Using user-based grouping significantly improves cache hit ratio by ensuring cached results align closely with user-specific request patterns.

\textbf{Access Count:} The total number of cache hits for each cached tool-calling result. This common caching metric helps ToolCaching quickly identify frequently requested entries (e.g., popular specific knowledge queries), directly informing prioritization in cache admission.

\textbf{Result Size:} The size of the cached tool-calling response. In LLM tool-calling scenarios, some requests (such as retrieving large datasets) produce much larger results than others. Tracking result size helps ToolCaching avoid allowing oversized entries to consume disproportionate cache space, ensuring a balanced trade-off between hit ratio and overall cache efficiency.

\textbf{System Latency:} The end-to-end time required to fulfill a tool-calling request, including both network transfer and tool execution delay. ToolCaching directly measures this latency within the cache system, allowing it to identify high-latency operations as prime candidates for caching. Prioritizing results with higher observed latency improves user-perceived responsiveness by minimizing repeated slow calls.

\emph{System-level indicators}, obtained via application-level measurement or low-level monitoring:  

\textbf{Observed Resource Cost:} The measured execution cost of each tool-calling request, either in terms of CPU consumption for local computations or billing price for external API calls. In LLM tool-calling, different tools have significantly varied resource costs; expensive local computations (e.g., personalized recommendation rankings) or costly third-party APIs (e.g., legal document retrieval) clearly justify higher caching value. ToolCaching leverages this feature to prioritize retention of results that yield substantial savings in computation or API billing costs.

For features that are difficult or costly to collect at the application layer, such as CPU usage of local APIs, we leverage lightweight system observability tools like eBPF (extended Berkeley Packet Filter)\cite{EBPF}, which is widely used for kernel-level performance tracing \cite{zhaiNRCACNonIntrusiveMicroservice2025,shenNetworkCentricDistributedTracing2023}. These allow per-request resource profiling with minimal runtime overhead. If needed, additional low-level metrics (e.g., memory usage, disk I/O, function calling) can also be captured through similar techniques to further inform caching decisions.

% A complete summary of the collected system features and their acquisition methods is provided in \autoref{tab:feature-collection}.

\section{Efficient Caching Management}\label{section:cachemanagement}
In this section, we discuss how to design efficient caching management policies based on the collected features.

\subsection{Cache Structure Design}
We begin by describing the design of the cache structure, which encompasses both cache key construction and value storage. This foundational design supports feature-aware admission and value-driven prioritization, laying the groundwork for the adaptive management policies discussed in subsequent sections.

\textbf{Cache Key Construction:} To ensure efficient and accurate cache lookups, we construct a unique key for each tool-calling request by combining the tool name with a serialized representation of its input parameters. The serialization process is designed to be order-invariant and supports complex, nested parameter structures (e.g., JSON objects), ensuring that logically equivalent requests yield identical keys. This composite string is then hashed to produce a compact, collision-resistant identifier. For example, a request to retrieve the weather for New York on May 1, 2024, would result in a key such as \texttt{hash(weather:{location=NewYork, date=2024-05-01})}. This approach enables fast lookups, accurate request distinction, and robust cache management in diverse LLM tool-calling scenarios.

\textbf{Cache Value Storage:} Each cache entry stores both the result of the tool call and its associated metadata. The metadata includes the relevant semantic and system features described in \autoref{section:feature_collection}, as well as a value score used to guide cache management decisions. The details of how this value score is computed are discussed later.

\subsection{Cacheability Analysis}
Before caching a tool-calling request, it is essential to determine whether the request is cacheable. We leverage a set of semantic features including request type, parameter category, and TTL to evaluate cacheability. Specifically, we adopt the following rules: (1) requests of type COMMAND are not cacheable; (2) requests with short TTL ($\leq 60$ seconds) are not cacheable.

Although these criteria are based on structured semantic features generated by the LLM, we deliberately avoid allowing the LLM itself to decide cacheability. Rule-based judgments over well-defined features ensure determinism, interpretability, and efficiency, whereas delegating this decision to the LLM could introduce uncertainty and potential errors. Overly complex or ambiguous LLM-driven analysis risks inconsistent cache states and degraded performance. Our approach therefore combines the LLM’s semantic extraction capability with transparent, rule-based cacheability assessment to maximize both expressiveness and reliability.

\subsection{Caching Value Model}
In addition to request access count, which is a standard metric in caching systems, we explicitly account for the overall cost of fulfilling each request in LLM tool-calling scenarios. This includes result size, system latency, and resource cost. By integrating these dimensions, our caching value model better reflects the real performance benefit of caching in complex, heterogeneous environments.

To ensure that these features are comparable and do not dominate the caching value due to scale differences, we apply min-max normalization to each metric. The normalization process rescales each feature $j$ of request $r_i$ as:

\begin{equation}
\text{NormFeature}_{i}^{(j)} = \frac{Feature_{i}^{(j)} - Feature_{min}^{(j)}}{Feature_{max}^{(j)} - Feature_{min}^{(j)}}
\end{equation}

where $Feature_{i}^{(j)}$ denotes the value of feature $j$ (e.g., latency, size, or cost) for request $r_i$, and $Feature_{min}^{(j)}$, $Feature_{max}^{(j)}$ are the minimum and maximum observed values of that feature, respectively.

Then, we can define the caching value $v_i$ of a request $r_i$ as follows:

% \begin{equation}
%     v_i = \alpha \cdot \text{NormLatency}_i 
%         + (1 - \alpha) \cdot \frac{\text{NormCost}_i}{\text{NormSize}_i}
% \label{eq:cache_value}
% \end{equation}
\begin{equation}
v_i = \lambda_1 \cdot \text{NormLatency}_i 
    + \lambda_2 \cdot \frac{\text{NormCost}_i}{\text{NormSize}_i}
    - \lambda_3 \cdot e^{-\frac{\text{TTL}_i}{\tau}}
\label{eq:cache_value}
\end{equation}
where $\lambda_1$, $\lambda_2$, and $\lambda_3$ are tunable weights controlling the relative importance of latency, cost-to-size ratio, and TTL-based risk in the caching value model. $\tau$ is a smoothing parameter that represents the average lifetime of cached entries. A smaller $\text{TTL}_i$ leads to a larger exponential term, indicating a higher risk of staleness, whereas a larger $\text{TTL}_i$ results in a smaller risk value, reflecting greater temporal stability.

Intuitively, the model prioritizes requests with high latency, large cost, and small result size (since they save system resources), while de-emphasizing those with short TTLs that are prone to rapid invalidation.
This balance between performance gain and temporal stability allows ToolCaching to cache entries that are both frequently accessed and sustainably valuable over time.

\subsection{Cache Admission}
Cache admission determines which requests enter the cache. In ToolCaching, we leverage the features collected in \autoref{section:feature_collection} to design a reinforcement learning-based admission policy, inspired by the CACA framework \cite{guanCACALearningbasedContentaware2019}, which adaptively adjusts to workload changes.

CACA first partitions requests into groups based on content features, then applies a reinforcement learning approach, which is specifically a multi-armed bandit algorithm, to dynamically tune admission policies for each group. By tracking frequency and hit ratio for each group, CACA selectively admits only the top-performing groups, focusing limited cache resources on requests with the greatest potential for reuse and system benefit. This approach is well-suited to the highly dynamic and diverse nature of LLM tool-calling workloads.

However, the original CACA only considers request frequency and hit ratio, overlooking fulfillment cost, a key factor in LLM tool-calling scenarios. We extend CACA to v-CACA by incorporating our caching value model into the reward function of the multi-armed bandit, so that both request patterns and execution cost guide cache admission.

Therefore, the cache admission policy in ToolCaching can be summarized as follows:

\textbf{Initial States:} In the initial phase, all cacheable tool-calling requests are admitted if space is available. For every request, features are collected and updated, especially upon cache hit or reuse.

\textbf{Feature Grouping:} 
To manage the diverse and dynamic nature of tool-calling requests, ToolCaching organizes requests into feature-based groups using a fixed three-level feature set: (1) \emph{tool type}, (2) \emph{parameter category}, and (3) \emph{user identity}. Requests are first partitioned by tool type, then (when needed) subdivided by parameter category, and finally by user identity to capture fine-grained locality patterns. The hierarchy is constructed \emph{adaptively}: a group is split only when its access frequency exceeds a threshold $T_1$ while its hit ratio is below $H^r$; otherwise it remains at its current level.

Each resulting group $g_i$ in the group list $\mathcal{G}$ maintains its own aggregate statistics including its cache hit ratio $H_i$ and the average caching value $V_i$.
This hierarchical feature grouping improves the stability of reinforcement learning decisions and allows the cache to respond adaptively to changing workloads. The full grouping logic, the corresponding hierarchy illustration and pseudo-code are provided in \autoref{appendix:admission}.

\textbf{Admission Control:} Now, we have a set of feature groups $\mathcal{G}$, each with its own average caching value $V_i$ and hit ratio $H_i$. The next step is to determine which feature groups should be admitted into the cache. 

This problem can be formulated as a multi-armed bandit problem, where each feature group $g_i$ is an arm, and the goal is to maximize the cumulative reward over time. Like CACA, we consider hit ratio $H_i$, node level $L_i$ and the number of times the group has been admitted to the cache $C_i$, but we also consider the average caching value $V_i$ of the requests in the group for the nature of tool-calling requests. Therefore, the reward function for each feature group $g_i$ is defined as follows:
\begin{equation}
    F_i = \frac{ \log(H_i + \delta_1) \cdot \log(L_i + \delta_2) \cdot \log(V_i + \delta_3) }{ \log(C_i + \delta_4) }
    \label{eq:reward}
\end{equation}  
where $\delta_1$, $\delta_2$, $\delta_3$, and $\delta_4$ are small constants to avoid division by zero and logarithm of zero.

To address this multi-armed bandit problem, we adopt UCB1 algorithm \cite{auerFinitetimeAnalysisMultiarmed2002}, which balances exploration and exploitation by selecting the arm with the highest upper confidence bound. The details of UCB1-based admission control are provided in \autoref{appendix:admission}.

\begin{figure*}[t]
    \centering
    \includegraphics[width=\textwidth]{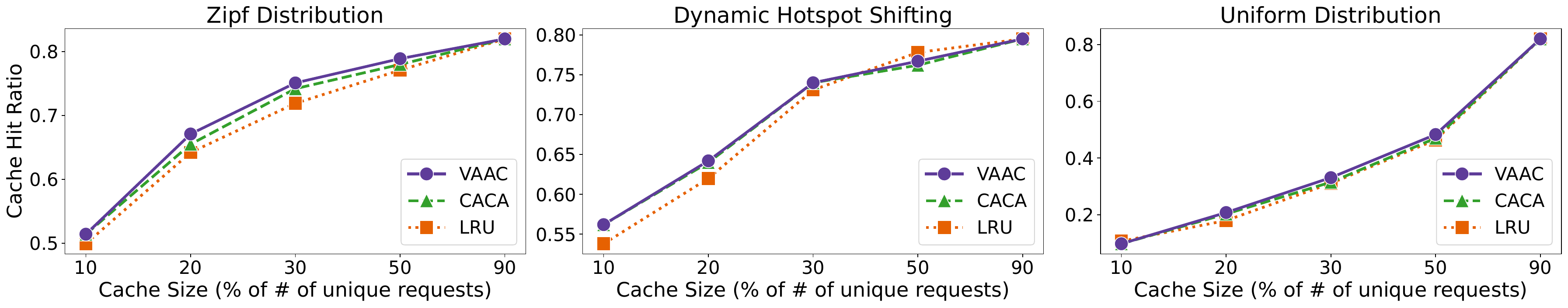}
    \caption{Cache hit ratio under different cache sizes and workloads}
    \label{fig:cache_hit_ratio}
\end{figure*}

\begin{figure*}[t]
    \centering
    \includegraphics[width=\textwidth]{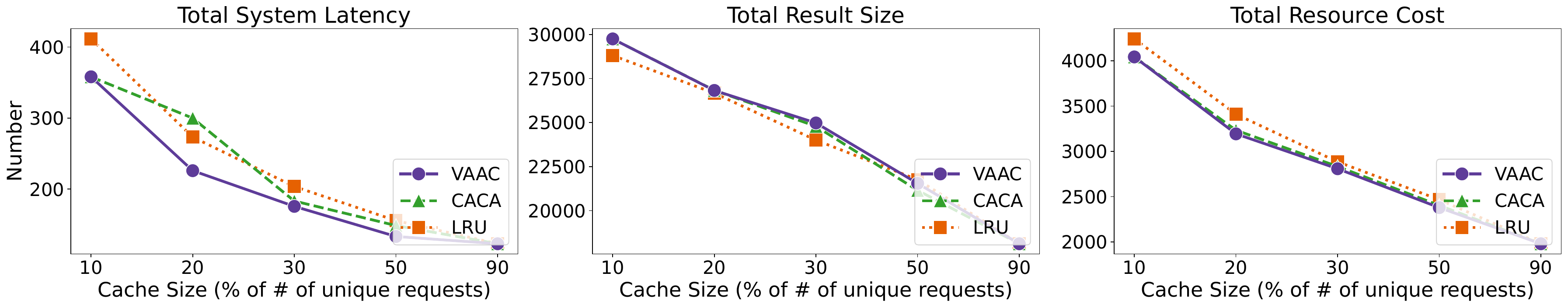}
    \caption{Total latency, result size and resource cost of requests under different cache sizes in Zipf distribution}
    \label{fig:zipf_detail}
\end{figure*}

\subsection{Cache Eviction}
Cache eviction determines which entries are removed when space is needed. In ToolCaching, we build on the classic LRU (Least Recently Used) policy and introduce v-LRU, a multi-factor eviction policy that jointly considers recency  and the caching value defined in Eq.~(\ref{eq:cache_value}).

The procedure of v-LRU is as follows:

\textbf{TTL Expiration:} Entries whose TTLs have explicitly expired are immediately invalidated. Although TTL-based staleness risk is already incorporated into the caching value $v_i$, explicit expiration handling ensures correctness and prevents outdated results from being accessed.

\textbf{Item Eviction:} When eviction is required, we identify the bottom 10\% of cache entries by recency (i.e., those least recently used). From this subset, we remove the entry with the lowest eviction score $e_i$, defined as:

\begin{equation}
    e_i = \log{(v_i+h_i+\delta_5)}
\end{equation}
where $v_i$ is the caching value, $h_i$ is the hit ratio, and $\delta_5$ is a small constant to ensure numerical stability.

Our strategy combines the caching value model with adaptive admission (v-CACA) and multi-factor eviction (v-LRU) to provide effective cache management for LLM tool-calling, together forming \textbf{Value-Aware Adaptive Caching (VAAC)} algorithm.

% \subsection{Cache Isolation}
% For tool calls that involve user-specific or sensitive content, it is crucial to ensure that cached results are isolated and not inadvertently shared across users. However, when access such tool calls, the specific \texttt{ACCESS\_ID} or \texttt{SECRET\_KEY} is in the parameters of the tool call, which makes impossible for cache data to be shared across users.

\section{Evaluation}\label{section:evaluation}

\begin{table}[b]
\centering
\caption{LLM semantic feature extraction accuracy on tool-calling requests.}
\renewcommand{\arraystretch}{1.2} % 调整行间距
\begin{tabular}{|l|cccc|}
\hline
\textbf{Features} & \textbf{Accuracy} & \textbf{Precision} & \textbf{Recall} & \textbf{F1} \\
\hline
Request Type & 0.980 & 0.968 & 1.000 & 0.984 \\
TTL & 0.920 & 0.864 & 0.864 & 0.864 \\
Parameter Category & 1 & 1 & 1 & 1 \\
\hline
\end{tabular}
\label{tab:llm_semantic_eval}
\end{table}

In this section, we evaluate the effectiveness of our feature collection and caching management algorithms in ToolCaching. We first assess the accuracy of the LLM in extracting semantic features from tool-calling requests, and then we evaluate the performance of our caching management algorithms using simulated workloads and public LLM tool-calling datasets. For the public dataset experiments, we implement our cache evaluation on top of the LLM Compiler framework. The cache server runs on a single machine with 16GB RAM and 8 CPUs, and the LLM used for semantic feature extraction and LLM Compiler is the cloud-based DeepSeek V3 \cite{liu2024deepseek,zhaoInsightsDeepSeekV3Scaling2025}.

\subsection{Semantic Feature Accuracy}
While semantic features are extracted using LLMs, a critical question arises: How accurate is the LLM in extracting the semantic features of tool-calling requests?

To answer this question, we conducted an evaluation of the LLM's semantic feature extraction capabilities on a set of tool-calling requests. Specifically, we extracted 50 tool-calling requests from the \texttt{BFCL\_v3\_simple} subset of the Berkeley Function Calling Leaderboard (BFCL) dataset \cite{patil2025bfcl}. Each request in our validation set was manually annotated with its corresponding request type (COMMAND or INFORMATIONAL) and an estimated TTL. These annotations serve as the ground truth for evaluating the accuracy of the LLM’s automatic classification and TTL prediction. To avoid confounding errors due to misparsed or malformed tool calls, we restrict our evaluation to only those cases where the LLM produces the correct tool call format. Therefore, since our system uses the first parameter of each tool call as the primary category, the correctness of parameter grouping is always $1$.

We use the same prompt template as shown in \autoref{appendix:prompt}, and we compute standard classification metrics—including accuracy, precision, recall, and F1-score—for each semantic attribute.

As shown in \autoref{tab:llm_semantic_eval}, the LLM achieves very high accuracy in extracting semantic features from tool-calling requests. The request type classification is nearly perfect, while TTL prediction also demonstrates strong accuracy. These results indicate that the LLM provides reliable semantic annotations for downstream caching management.

\begin{table}[b]
\centering
\caption{Comparison of cache hit ratio and latency with/without user-based grouping.}
\renewcommand{\arraystretch}{1.2} % 调整行间距
\begin{tabular}{|l|c|c|c|}
\hline
Cache Size             & 10\%            & 20\%            & 30\%            \\
\hline
With User    & 18.8 / 563.28 & 32.2 / 465.60 & 52.1 / 346.32 \\
Without User & 15.5 / 568.88 & 30.2 / 501.04 & 51.2 / 363.62 \\
\hline
\end{tabular}
\label{tab:user_cache_summary}
\end{table}

\begin{table*}[t]
    \centering
    \caption{Effectiveness of ToolCaching in LLM Compiler}
    \renewcommand{\arraystretch}{1.2}
    \begin{tabular}{|l|c|c|c|c|}
    \hline
    % \textbf{Dataset} & \textbf{Cache Size (of calls' \#)} & \textbf{Accuracy} & \textbf{Cache Hit Ratio} & \textbf{Latency(s)} & \textbf{Improvement} \\
    \textbf{Dataset} & \textbf{Cache Size (of calls' \#)} & \textbf{Cache Hit Ratio} & \textbf{Latency(s)} & \textbf{Improvement} \\

    \hline
    \multirow{4}{*}{Movie Recommendation} 
        % & 0\%   & 0.88 & ---   & 16.2  & ---   \\
        & 0\%   & ---   & 16.2  & ---   \\
        & 20\%  & 0.3781 & 13.81 & 1.14$\times$ \\
        & 50\%  & 0.503  & 11.88 & 1.26$\times$ \\
        & 100\% & 0.514  & 10.7  & 1.34$\times$ \\
    \hline
    \multirow{4}{*}{ParallelQA} 
        % & 0\%   & 0.82 & ---   & 15.9  & ---   \\
        & 0\%   & ---   & 15.9  & ---   \\
        & 20\%  & 0.232  & 13.08 & 1.17$\times$ \\
        & 50\%  & 0.265  & 12.99 & 1.18$\times$ \\
        & 100\% & 0.2866 & 12.18 & 1.23$\times$ \\
    \hline
    \end{tabular}
    \label{tab:toolcaching_effectiveness}
\end{table*}

\subsection{Caching Management}
Due to the lack of publicly available datasets that accurately reflect real-world LLM tool-calling workload patterns, we construct a synthetic dataset to evaluate the performance of our caching management algorithms.

To comprehensively evaluate the effectiveness of our caching management algorithms, we construct three types of synthetic workloads, each consisting of 1,000 requests covering 6 different tools, which are designed to cover a diverse range of parameter types, response sizes, TTLs, and resource costs, ensuring broad coverage of typical tool-calling scenarios:
\begin{itemize}
    \item \textbf{Zipf distribution:} Requests are sampled according to a Zipf distribution ($\alpha = 1.1$), mimicking real-world workload locality~\cite{abolhassaniOptimalLoadSplittingDistributedCaching2023, dengFundamentalStructureOptimal2022, guanCACALearningbasedContentaware2019, breslauWebCachingZipflike1999}.
    \item \textbf{Dynamic hotspot shifting:} The workload periodically switches between different hotspot regions. In each phase, request frequencies within the active hotspot follow a Zipf distribution, while other requests are distributed among the remaining tools.
    \item \textbf{Uniform distribution:} Requests are generated uniformly at random across all tool and parameter combinations, serving as a baseline to evaluate cache policy performance under minimal locality.
\end{itemize}

We implement the cache server for simulated workloads using Python with $\sim$1400 lines of code and evaluate the performance of cache hit ratio under 5 different cache sizes: 10\% of all unique requests, 20\%, 35\%, 50\% and 90\%. We set $\lambda_1=0.8$, $\lambda_2=0.2$, $\lambda_3=0.2$, $T_1=20$ and $H_r=0.5$.

For each cache size, we compare the performance of our VAAC algorithm against two baseline algorithms: CACA (admits requests without consideration of value and LRU eviction policy) and LRU (admits all requests into the cache). The results are shown in \autoref{fig:cache_hit_ratio}. As depicted, VAAC consistently outperforms both CACA and LRU across most cache sizes and workloads by up to 11\%. This demonstrates the effectiveness of our caching value model and multi-factor eviction policy in optimizing cache performance. 

Specifically, we present the total system latency, result size and resource cost of requests under different cache sizes in the Zipf distribution workload in \autoref{fig:zipf_detail}. As shown, VAAC achieves the lowest total latency and resource cost compared to CACA and LRU with up to 17.3\% and 6.4\% reduction, respectively. For the total result size, due to our cache value model penalizing large result sizes in \autoref{eq:cache_value}, VAAC achieves the highest result because high-value requests are more likely to be refused and evicted, which is consistent with our design goal of maximizing cache utility.

\subsection{Multi-user Cache Effectiveness}
As multi-user scenarios are common in practical LLM applications, we evaluate ToolCaching under realistic settings by simulating a workload with 10 users, where each user issues tool-calling requests with partially overlapping interests. We compare two cache management strategies: (1) global grouping (no user separation) and (2) user-based grouping (our approach). Results in \autoref{tab:user_cache_summary} show that user-based grouping achieves up to 21.3\% higher cache hit ratio and 7.1\% lower average latency. This demonstrates that ToolCaching effectively adapts to user-specific access patterns, improving user-perceived performance.

\subsection{End-to-End Evaluation in LLM Tool-calling Systems}
The development of tool-calling has enabled LLMs to solve increasingly complex problems by selecting and coordinating multiple functions based on context. To further support such applications, frameworks like LLM Compiler \cite{kimLlmCompilerParallel2024} provide optimization for multi-step reasoning and tool orchestration.

To evaluate the real-world effectiveness of ToolCaching in advanced LLM tool-calling scenarios, we integrate our caching management system into the LLM Compiler framework and conduct an end-to-end evaluation. We use the following dataset provided by LLM Compiler:

\begin{itemize}
    % \item \textbf{HotpotQA:} HotpotQA is a benchmark for multi-hop reasoning. We use the comparison development set, which consists of 1,500 questions requiring comparison between two entities. This setting corresponds to a 2-way embarrassingly parallel execution pattern, as each comparison can be resolved independently.
    \item \textbf{Movie Recommendation:} This dataset contains 500 examples, each asking for the most similar movie among four options compared to a reference set of four movies. The task exhibits an 8-way embarrassingly parallel execution pattern, where each candidate can be evaluated separately.
    \item \textbf{ParallelQA:} ParallelQA is a custom benchmark for evaluating tool-calling scenarios with complex dependencies. It includes 113 math-related questions about factual attributes, where each task requires sequential use of two tools (such as search followed by math) with the second tool’s input depending on the first tool’s output. All required information is contained within the first paragraph of relevant Wikipedia articles.
\end{itemize}

We compare the performance of ToolCaching with the baseline LLM Compiler framework without caching. The results are summarized in \autoref{tab:toolcaching_effectiveness}. As shown in the table, ToolCaching consistently reduces end-to-end latency as the cache size increases. In the Movie Recommendation dataset, ToolCaching achieves up to a 34\% reduction in latency. Similarly, for ParallelQA, the latency is reduced by up to 23\%. These results demonstrate that ToolCaching can effectively reduce redundant tool calls and significantly improve system efficiency in complex LLM tool-calling scenarios.

% It is worth noting that, in all three datasets, at most two tools are involved, and each tool has only a single parameter. In addition, these tools perform searches by querying Wikipedia and only extract information from the first paragraph, so their size and cost are essentially similar. Under such circumstances, our grouping strategy becomes meaningless, and value estimation also loses effectiveness due to the nearly identical latency and overhead, making the admission decisions almost random. Similarly, v-LRU degenerates into random selection among the last 10\% of LRU items. Therefore, these experiments are only intended to demonstrate the effectiveness of caching for tool-calling scenarios, and cannot fully demonstrate the effectiveness of the v-CACA algorithm.

\subsection{Overhead}
We also measure the overhead of ToolCaching in the scenario of LLM Compiler. The CPU overhead of ToolCaching is $\sim 15\%$ and the memory overhead is $\sim 10\%$ compared to the baseline LLM Compiler framework without caching. The CPU overhead mainly comes from the calculation of VAAC algorithm, while the memory overhead is due to the storage of cached entries and their associated metadata. These overheads are acceptable given the significant performance improvements achieved through caching.

\section{Conclusion}
In this paper, we present ToolCaching, an efficient caching framework for LLM tool-calling systems. At its core, ToolCaching employs VAAC algorithm, which integrates semantic and system-level features, dynamically partitions requests into groups, and utilizes a bandit-based admission policy together with value-aware LRU eviction. Through extensive experiments, we demonstrate that ToolCaching with VAAC achieves higher cache hit ratios and lower latency than conventional static and frequency-based policies. Our results further show that ToolCaching adapts effectively to dynamic workloads and varying result validity, underscoring the value of caching in complex LLM environments.

%%
%% The next two lines define the bibliography style to be used, and
%% the bibliography file.
\bibliographystyle{ACM-Reference-Format}
\bibliography{reference}{}

\appendix

\section{Semantic Feature Extraction Details}\label{appendix:prompt}
\begin{figure}[t]
    \centering
    \includegraphics[width=0.48\textwidth]{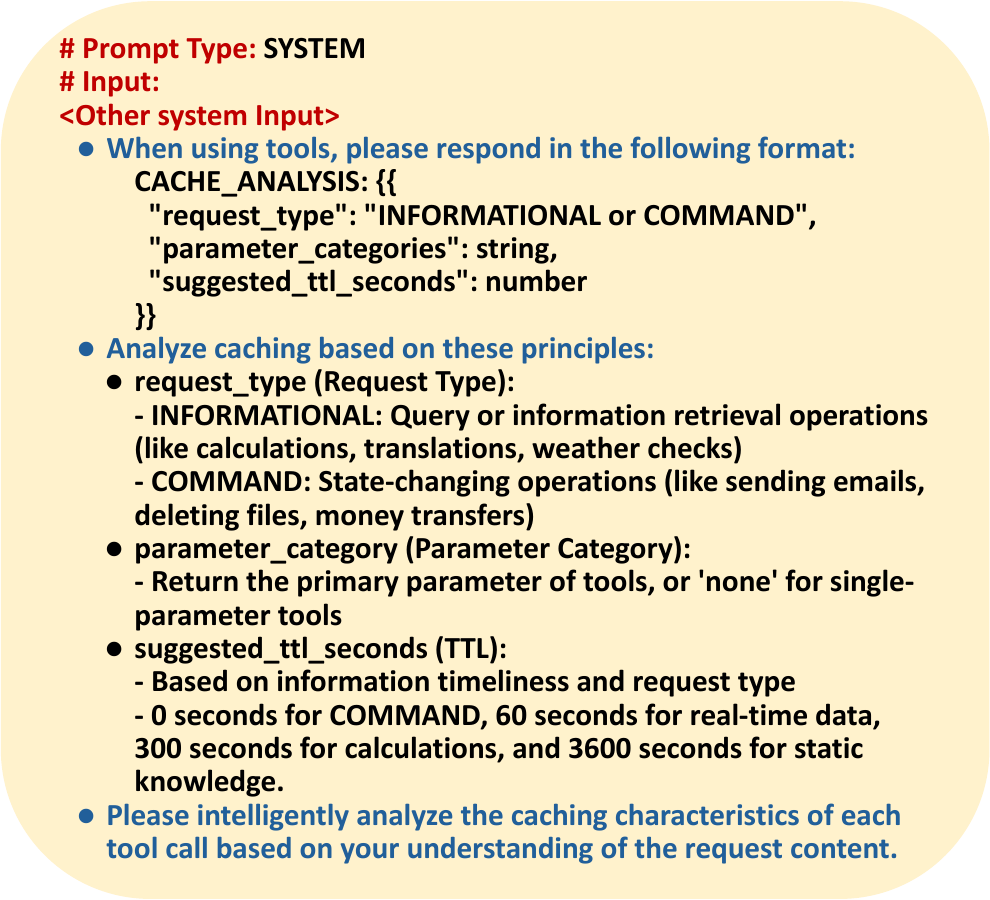}
    \caption{Prompt template for LLM semantic feature extraction.}
    \label{fig:prompt}
\end{figure}

We provide the detailed prompt template used for LLM-based feature extraction in \autoref{fig:prompt}.

\section{Cache Admission Details}\label{appendix:admission}
\subsection{Hierarchical Feature Grouping}
We adopt a hierarchical and adaptive grouping strategy. Requests are initially partitioned by tool type. For each tool group, we continuously track aggregate statistics such as access count and cache hit ratio. If a group has high frequency (request access count $\geq T_1$) but low hit ratio ($\leq H^r$), which suggests substantial diversity, we further subdivide by parameter category. If parameter-based grouping is still insufficient to capture homogeneity, we refine the grouping by user ID. This is especially useful in multi-user environments where access patterns can differ significantly.

\begin{figure}[t]
    \centering
    \includegraphics[width=0.48\textwidth]{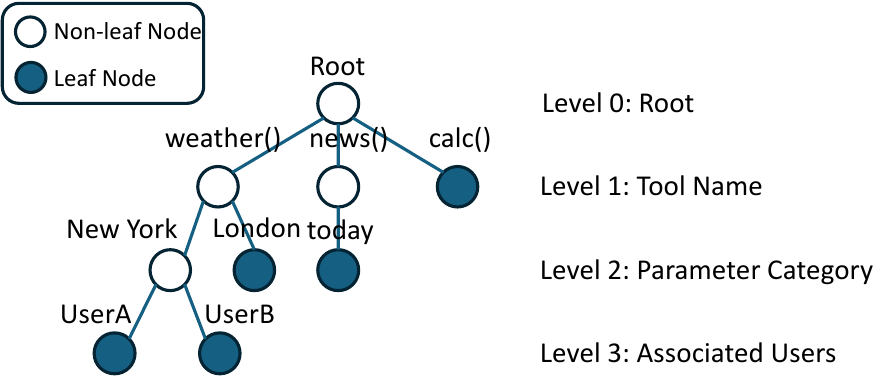}
    \caption{Hierarchical feature grouping for tool-calling requests.}
    \label{fig:tree}
\end{figure}

Through this recursive process, requests are organized into a tree structure (\autoref{fig:tree}), with each layer corresponding to a specific feature dimension. The leaf nodes represent homogeneous request clusters managed by the cache. For example, a leaf like \texttt{root/weather()/New York/UserA} represents a group for "weather" tool with the parameter "New York" and user "UserA". To prevent over-fragmentation, groups with too few requests are merged with their parent group.

This dynamic refinement process, guided by frequency and hit ratio, allows the system to maximize cache utility by adaptively aligning group granularity with real-world workload patterns. At the same time, it helps control system complexity.

After feature grouping, a list of feature groups is formed: 
\begin{equation}
    \mathcal{G} = \{g_1, g_2, \ldots, g_n\}
\end{equation}
where each group $g_i$ contains requests with similar features. For each group $g_i$, we maintain a set of statistics, including the average caching value of the requests $V_{i} = \frac{1}{|g_i|} \sum_{r_j \in g_i} v_j$, and the cache hit ratio $H_i$. 

If there are too few requests ($< S_{min}$) in a group, we merge it with its parent group to reduce the number of groups. Also, the group will be periodically reset after $B$ requests to adapt to changing workloads. 

We provide the detailed algorithm for periodic hierarchical feature grouping in \autoref{alg:periodic_feature_grouping}.
\begin{algorithm}[h]
    \caption{Periodic Hierarchical Feature Grouping for Tool-calling Requests}
    \label{alg:periodic_feature_grouping}
    \begin{algorithmic}[1]
    \REQUIRE Request stream $\mathcal{R}$, feature dimensions $\mathcal{F}$, thresholds $T_1$ (freq), $H^r$ (hit ratio), min group size $S_{\min}$, regroup interval $B$
    \ENSURE Leaf feature groups $\mathcal{G}$

    \STATE Initialize $\mathcal{G} \leftarrow \emptyset$
    \STATE Initialize request buffer $\mathcal{R}_{\text{all}} \leftarrow \emptyset$, counter $cnt \leftarrow 0$
    \STATE \textbf{procedure:} \texttt{PeriodicGrouping}($\mathcal{F}$, $B$)
    \begin{ALC@g}
        \FOR{each incoming request $r$}
            \STATE Add $r$ to $\mathcal{R}_{\text{all}}$, $cnt \gets cnt + 1$
            \IF{$cnt$ mod $B == 0$}
                \STATE $\mathcal{G} \leftarrow \emptyset$
                \STATE \texttt{HierarchicalGrouping}($\mathcal{R}_{\text{all}}$, $\mathcal{F}$, 1)
                \STATE // Now $\mathcal{G}$ stores the latest grouping result
            \ENDIF
        \ENDFOR
    \end{ALC@g}

    \STATE \textbf{procedure:} \texttt{HierarchicalGrouping}($\mathcal{R}$, $\mathcal{F}$, level)
    \begin{ALC@g}
        \IF{level $> |\mathcal{F}|$ \textbf{or} $|\mathcal{R}| < S_{\min}$}
            \STATE Add $\mathcal{R}$ as a group to $\mathcal{G}$
            \STATE \textbf{return}
        \ENDIF
        \STATE Partition $\mathcal{R}$ into subgroups $\{\mathcal{R}_1, \mathcal{R}_2, \ldots\}$ by feature $\mathcal{F}[$level$]$
        \FOR{each subgroup $\mathcal{R}_j$}
            \STATE Compute access count $f_j$, hit ratio $H_j$
            \IF{$f_j \geq T_1$ \textbf{and} $H_j \leq H^r$}
                \STATE \texttt{HierarchicalGrouping}($\mathcal{R}_j$, $\mathcal{F}$, level + 1)
            \ELSIF{$|\mathcal{R}_j| < S_{\min}$}
                \STATE Add $\mathcal{R}_j$ to parent group (skip splitting)
            \ELSE
                \STATE Add $\mathcal{R}_j$ as a leaf group to $\mathcal{G}$
            \ENDIF
        \ENDFOR
    \end{ALC@g}

    \STATE \textbf{output:} Latest leaf feature groups $\mathcal{G}$ (regrouped every $B$ requests)
    \end{algorithmic}
\end{algorithm}

\begin{algorithm}[h]
    \caption{Cache Admission via UCB1-based Algorithm}
    \label{alg:cache_admission}
    \begin{algorithmic}[1]
    \REQUIRE Leaf feature groups $\mathcal{G} = \{g_1, g_2, \ldots, g_n\}$ from hierarchical partitioning; current statistics for each $g_i$
    \ENSURE Admission decisions for each decision round

    \STATE \textbf{procedure:} \texttt{UCB1-Admission}$(\mathcal{G})$
    \begin{ALC@g}
        \FOR{each decision round $t$}
            \FORALL{feature group $g_i \in \mathcal{G}$}
                \STATE Update statistics for $g_i$: hit ratio $H_i$, node level $L_i$, admission count $C_i$, avg. caching value $V_i$, selection count $N_i$
                \STATE Compute reward $F_i$ using Eq.~(\ref{eq:reward})
                \STATE Compute $UCB_i = F_i + c \sqrt{\frac{\ln t}{N_i}}$ using Eq.~(\ref{eq:ucb1})
            \ENDFOR
            \STATE Select $g_{i^*} = \arg\max_{g_i} UCB_i$
            \STATE Admit requests from $g_{i^*}$ into the cache
            \STATE Update $C_{i^*} \gets C_{i^*} + 1$, $N_{i^*} \gets N_{i^*} + 1$
        \ENDFOR
    \end{ALC@g}
    \STATE \textbf{end procedure}
    \end{algorithmic}
\end{algorithm}

\subsection{UCB1-based Admission Control}
We adopt UCB1 algorithm \cite{auerFinitetimeAnalysisMultiarmed2002} to address the multi-armed bandit problem in cache admission. At each decision round, we calculate the reward $F_i$ for each group according to \autoref{eq:reward}, and select the group with the highest UCB value:

\begin{equation}
    UCB_i = F_i + c \sqrt{\frac{\ln t}{N_i}}
\label{eq:ucb1}
\end{equation}
where $N_i$ is the number of times group $i$ has been selected, $t$ is the current round, and $c$ is an exploration parameter.

The psudo-code of UCB1-based admission control is shown in \autoref{alg:cache_admission}.

\section{Discussion on Limitations and Future Work}
\subsection{Limitations}
Although ToolCaching demonstrates substantial performance gains in LLM tool-calling scenarios, it primarily targets \textbf{INFORMATIONAL} tool calls, where the result of a call does not alter the system state. 
However, with the increasing adoption of the \textbf{Model Context Protocol (MCP)}~\cite{houModelContextProtocol2025}, LLMs are becoming capable of invoking \textbf{COMMAND}-type tools that perform state-changing or irreversible operations. 

As discussed in \autoref{section:feature_collection}, applying traditional caching strategies to such calls may lead to harmful side effects. Therefore, future caching frameworks must explicitly account for the semantics and side-effect characteristics of COMMAND calls, adopting fundamentally different mechanisms from those used for INFORMATIONAL requests.

\subsection{Future Work}

For future work, we plan to extend ToolCaching to support more diverse tool-calling patterns, multi-tenant scenarios, and integration with multi-agent frameworks, particularly exploring caching opportunities across inter-agent tool-calling patterns.
Also, we aim to explore how caching can be safely and effectively applied to \textbf{COMMAND}-type tool calls, thereby enhancing performance in MCP scenarios.

\end{document}